\documentclass[journal]{IEEEtran}
\usepackage{color}
\usepackage{graphicx}
\usepackage{subfigure}
\usepackage{amsmath}
\usepackage{amsthm}
\usepackage{float}
\usepackage{epsfig}
\usepackage{cite}
\usepackage{url}

\theoremstyle{remark}

\theoremstyle{definition}

\newfloat{problem}{htb}{loM}
\floatname{problem}{Problem}
 
\usepackage{booktabs}
\usepackage{multirow}
\usepackage{mathrsfs}
\usepackage{amsfonts}
\usepackage{algorithm,algorithmic}
\usepackage{paralist}
\usepackage{tabularx}
\usepackage{amssymb}
\usepackage{subfigure}
\usepackage{graphicx}
\usepackage{float}

\ifCLASSINFOpdf
\else
\fi
\begin{document}
\graphicspath{{images/}}

\title{EC-SAGINs: Edge Computing-enhanced Space-Air-Ground Integrated Networks for Internet of Vehicles}

\author{Shuai~Yu, %~\IEEEmembership{Member,~IEEE,}
        Xiaowen~Gong, %~\IEEEmembership{Fellow,~OSA,}
        Qian Shi,
        Xiaofei Wang
        and~Xu~Chen %~\IEEEmembership{Life~Fellow,~IEEE}% <-this % stops a space
\thanks{Shuai~Yu, Qian~Shi and Xu~Chen are with Sun Yat-sen University, Guangzhou 510275, China (Emails: \{yushuai, shixi5, chenxu35\}@mail.sysu.edu.cn).}% <-this % stops a space
\thanks{Xiaowen~Gong is with Auburn University, Auburn, AL, USA (E-mail: xgong@auburn.edu). }% <-this % stops a space
\thanks{Xiaofei Wang is with Tianjin University, China (e-mail: xiaofeiwang@tju.edu.cn).}
%\thanks{Corresponding Author: Qian Shi.}
%\thanks{Copyright (c) 20xx IEEE. Personal use of this material is permitted. However, permission to use this material for any other purposes must be obtained from the IEEE by sending a request to pubs-permissions@ieee.org.}
}

\maketitle

\begin{abstract}
Edge computing-enhanced Internet of Vehicles (EC-IoV) enables ubiquitous data processing and content sharing among vehicles and terrestrial edge computing (TEC) infrastructures (e.g., 5G base stations and roadside units) with little or no human intervention, plays a key role in the intelligent transportation systems.
However, EC-IoV is heavily dependent on the connections and interactions between vehicles and TEC infrastructures, thus will break down in some remote areas where TEC infrastructures are unavailable (e.g., desert, isolated islands and disaster-stricken areas).
Driven by the ubiquitous connections and global-area coverage, space-air-ground integrated networks (SAGINs) efficiently support seamless coverage and efficient resource management, represent the next frontier for edge computing.
In light of this, we first review the state-of-the-art edge computing research for SAGINs in this article.
After discussing several existing orbital and aerial edge computing architectures, we propose a framework of edge computing-enabled space-air-ground integrated networks (EC-SAGINs) to support various IoV services for the vehicles in remote areas.
The main objective of the framework is to minimize the task completion time and satellite resource usage.
To this end, a pre-classification scheme is presented to reduce the size of action space, and a deep imitation learning (DIL) driven offloading and caching algorithm is proposed to achieve real-time decision making.
Simulation results show the effectiveness of our proposed scheme.
At last, we also discuss some technology challenges and future directions.
\end{abstract}


% \fi
\IEEEpeerreviewmaketitle

\section{Introduction}
Evolving from vehicular ad hoc networks (VANETs), Internet of Vehicles (IoV) is a state-of-the-art theme of Internet of Things (IoT) for supporting various information services, such as dynamic road information services, vehicle automatic control and intelligent transportation systems~\cite{9088328}.
One of the key challenges of IoV is its limited capacity for handling and processing the massive data that is collected by vehicles and other sensors around the environment.
%±ßÔµ¼ÆËãÎÊÌâ
In view of this, edge computing~\cite{8736011} is considered to be an advanced computing paradigm with data processed at network edge.
According to this paradigm, computation and storage resources are placed in close proximity to vehicles.
Thus, the bandwidth demands on network as well as computation and communication latencies can be reduced.
Although current studies on edge computing-enhanced Internet of Vehicles (EC-IoV)~\cite{8786080,8318667,8770298} have achieved great advances in providing real-time and energy-efficient road services for vehicles, they still face the challenges of limited network coverage and scarce network resources.
On the one hand, terrestrial EC-IoV cannot provide fair and high quality services for all the vehicles around the world, since it is impossible for them to achieve global and seamless coverage.
Note that, over 50 percent of the world, especially for some complex terrains areas, such as mountains, oceans and isolated islands, still lacks of network coverage.
On the other hand, terrestrial EC-IoV infrastructures are vulnerable to natural disasters (e.g., earthquakes and flood).
Thus, vehicles in disaster-stricken areas may lose network connections.

%¸²¸ÇÎÊÌâ
Driven by the ubiquitous connections and global-area coverage of space-air-ground integrated networks (SAGINs), recent years have seen a paradigm shift in edge computing, from terrestrial edge computing toward aerial/orbital edge computing, especially in the remote areas.
The main feature of aerial/orbital edge computing is to push terrestrial edge computing facilities to the sky (e.g., in satellite and unmanned aerial vehicle) so as to enable ubiquitous, broadband and reliable cloud computing services.
%During the past years, space air ground integrated networks (SAGINs) which consist of orbital networks, aerial networks and terrestrial infrastructures have been regarded as novel network architectures to provide global-range and seamless networking services~\cite{8994207}.
In addition, the advancement of satellite technologies (e.g., manufacturing and laser transmission) makes satellites, especially low Earth orbit (LEO) satellites, much more economical and high data rate.
Thus, SAGINs represent the next frontier for edge computing.
Specifically, applying edge computing techniques to the orbital segment of SAGINs can greatly extend the coverage of terrestrial EC-IoV.
The aerial segment of SAGINs has the advantages of low-cost and flexible networking for TEC.
Thus, vehicles around the world can leverage the global coverage, flexibility and high efficiency edge computing services of SAGINs.
Moreover, the edge computing enhanced SAGINs can also be applied in various practical fields that require real-time data sensing and processing, such as weather monitoring, Earth observation, disaster relief, precision agriculture and smart cities.

However, integrating edge computing into SAGINs still faces numerous challenges.
First, delay and bandwidth are the main challenges for SAGINs, especially for geostationary Earth orbit (GEO) and medium Earth orbit (MEO) satellite systems.
The large one way delay of GEO and MEO satellites degrades the performance of real-time applications.
Then, the energy supply for satellites and high altitude platforms (HAPs, such as unmanned aerial vehicles, airships, and balloons) is limited.
For example, due to the small size and limits solar panel surface area of LEO satellites, it is hard for the satellites to run some high energy consumption applications, such as deep learning algorithm.
At last, most of today's IoV applications involve complex transmission (e.g., movie downloads), processing (e.g., video analysis and speech recognition) and caching (content and service caching) process.
Thus, low latency task processing schemes and efficient resource management strategies are required for the edge computing enhanced SAGINs.

In this paper, we study the issue of how to efficiently integrate edge computing into SAGINs, in order to support various IoV services.
The main contributions of this article are summarised as follows:
\begin{itemize}
\item We first review state-of-the-art edge computing techniques in SAGINs, and summarize some available architectures, use cases, advantages and challenges for them.
To the best of our knowledge, this is the first paper to survey the edge computing in SAGINs, since existing articles focus on either specific characteristics of edge computing, or the architecture of SAGINs, neglecting the integration of the above network segments.
We also propose a framework of edge computing enabled space air ground integrated networks (EC-SAGINs), and formulate the fine-grained offloading and caching problem as a multi-label classification process.
\item Next, we present a pre-classification scheme to reduce the action space, and propose an offline deep imitation learning (DIL) based offloading and caching algorithm to achieve real-time decision making.
Current studies on deep learning based intelligent offloading or caching schemes generally concentrate on the deep reinforcement learning (DRL) approaches~\cite{8941121,8434316,9026935,8879573}.
However, the online training manner of DRL is usually computation intensive and high energy consuming, thus can not be applied in the LEO satellites with limited energy supply.
At last, we also discuss potential directions and advantages for applying edge computing into SAGINs.
\end{itemize}

The rest of this article is organized as follows.
In Section~\ref{sec:Related_Works}, we introduce the related works most relevant to this article.
Section~\ref{sec:review} reviews state-of-the-art orbital edge computing and aerial edge computing technologies, respectively.
The framework of EC-SAGINs is proposed in Section~\ref{sec:EC-SAGIN}.
Section~\ref{sec:System-Model} presents the system of EC-SAGINs,
We formulate the optimization problem in Section~\ref{sec:Problem-Formulation}, followed by a description of our proposed schemes in Section~\ref{sec:Algorithm_Design}.
Simulation results are presented in Section~\ref{sec:performance_evaluation}.
We further discuss directions and advantages of edge computing for SAGINs in Section~\ref{sec:extensions}.
Finally, we conclude the article in Section~\ref{sec:Conclusion}.

\section{Related Works}{\label{sec:Related_Works}}
In this section, we will survey traditional terrestrial edge computing (TEC) and state-of-the-art space-air-ground integrated networks (SAGINs), respectively.

%Recent years, edge computing~\cite{8736011} is a state-of-the-art computing paradigm by moving the cloud computing services and resources closer to end users.
%Some well known terrestrial edge computing technologies, such as multi-access edge computing (MEC)~\cite{ETSIMEC} and fog computing~\cite{8100873} are proposed and applied into terrestrial wireless communications systems.
%MEC is proposed by European Telecommunications Standards Institute (ETSI) Industry Specification Group (ISG), in order to transfer computation and storage capacity from remote cloud to the edge of 5G wireless networks.
%Thus, many benefits can be obtained by end users, such as high data rates, saved bandwidth resources and improved the quality of service (QoS) for some real-time services.
%Specifically, MEC provides computation offloading and content/service caching functions to realize the above advantages~\cite{10.1145/3362031}.
%For the end users running computation-intensive tasks (e.g., virtual reality and deep learning algorithm), they can dynamically offload the computation-intensive part of their tasks to edge servers instead of remote cloud servers (i.e., computation offloading).
%On the other hand, the users can also benefit from the fact that their required contents/services have already been cached in their nearby edge servers (i.e., content/service caching).

\subsection{Edge Computing}
Edge computing~\cite{8736011} is a state-of-the-art computing paradigm by moving the cloud computing services and resources closer to end users, in order to achieve high data rates, saved bandwidth resources and improved the quality of service (QoS) for real-time services.
Recent years, some state-of-the-art technologies are integrated into edge computing networks, in order to achieve efficient resource allocation.
For example, authors in~\cite{9060868} survey Federated Learning (FL) in edge computing networks.
They show that FL can serve as an enabling technology in managing heterogeneous edge resources.
Blockchain based resource allocation is studied in~\cite{8624417}.
Note that, the resources consumed by blockchain are non-negligible.
Current studies on edge computing-enhanced Internet of Vehicles (EC-IoV) generally concentrate on resource management or computation offloading in the terrestrial (especially the urban) scenarios.
For example, authors in~\cite{8786080} propose an energy-efficient scheduling method for the MEC-enabled Internet of Vehicles.
The main objective is to minimize the energy consumption of roadside units under task completion time constraints.
Authors in~\cite{8770298} investigate energy efficient task offloading in intelligent IoV scenario.
They assume that both parked vehicles and moving vehicles are fog nodes, and propose a three-layer task offloading framework.

Although TEC can provide real-time services and high data rate to end users, its network coverage in rural and remote areas (e.g., mountain areas and highways) is limited.
In addition, it is very expensive to deploy expensive edge computing infrastructure in those sparsely areas.
Take autonomous vehicles technology as an example, it requires edge servers to provide seamless and real-time services to moving vehicles.
However, the coverage diameter of an edge server (e.g., small cell base station) is usually less than 300 m.
Consequently, moving vehicles will experience highly frequent handovers in TEC networks.
More seriously, for the vehicles in rural or remote highways, they may even lose network connections.

\subsection{Space-Air-Ground Integrated Networks}
%³ÐÉÏÆôϼò½é
Due to the ubiquitous and global-area coverage of SAGINs, they have been becoming promising techniques evolving to 6G and attracted much attention during the past years~\cite{8368236}.
SAGINs are hierarchical networks which integrate orbital networks in the space, aerial networks in the air and terrestrial networks in the ground.
The above networks can work independently or inter-operationally.
%orbital network
The orbital segment is composed of multi-layered satellites and the corresponding terrestrial infrastructures (e.g., ground stations and data center), provides ubiquitous and wide-area communication for navigation, Earth observation, emergency relief, space exploration, etc.
Specifically, the multi-layered satellite system is composed of the satellites in different orbits, i.e., Geostationary Earth Orbit (GEO) satellites, Medium Earth Orbit (MEO) satellites, and Low Earth Orbit (LEO) satellites.
The orbital segment has the following advantages: i) provides seamless and flexible network connectivity to large areas, such as islands, rural and isolated mountainous areas; ii) improves the network capacity by leveraging multi-cast and broadcast techniques.
%aerial network
The aerial segment consists of high altitude platforms (HAPs), offers broadband wireless services complementing the terrestrial network.
Note that the HAPs can be partially controlled by terrestrial controller on demand, in order to provide supplemental communication capacity for hotspot areas and network coverage for isolated areas.
The aerial segment has the advantages of low cost, large coverage and easy deployment.
%terrestrial network
The terrestrial segment is mainly composed of ground communication systems such as 5G cellular networks, MANETs, WiMAX and WLANs.
Densely deployed terrestrial network can provide highest throughput and rich resources to end users.
However, the network coverage in remote areas is quite limited, and the network facilities are vulnerable to natural disasters.
SAGINs are also facing many unprecedented challenges such as high bit error rate, long propagation latency, and unstable connectivity.
Thus, how to efficiently integrate edge computing functions into SAGINs should be carefully considered.

\begin{footnotesize}
\begin{table*}[tp]
\caption{Comparison of different edge computing technologies for SAGINs.}
\label{tab:Comparison}
\centering
\renewcommand\arraystretch{1}
\begin{tabular}{|m{13mm}<{\centering}|m{15mm}<{\centering}|m{20mm}<{\centering}|m{10mm}<{\centering}|m{18mm}<{\centering}|m{20mm}<{\centering}|m{20mm}<{\centering}|m{28mm}<{\centering}|}\hline
Segment                        &Architectures                                           &Hight                          &Coverage &One way delay                           &Use cases  &Advantages  &Challenges\\\hline
Orbital edge computing         &OEC~\cite{OEC}, SMEC~\cite{8610431}.                    & 160-2000km (LEO), 2000-35786km (MEO), 35786km (GEO).   &Global.   &270ms (GEO), 110ms (MEO), less than 40ms (LEO).    &Earth observation~\cite{Tokyo_Tech}, vehicular network~\cite{7981533}.    & Global coverage.    & Vulnerable to radiation, long propagation latencies, limited energy supply.\\\hline
Aerial edge computing          &AGMEN~\cite{8436041}, UAV-Edge-Cloud~\cite{8675170}.     & 17-30km (HAP).                 &Wide.     &Medium.                                  &Smart cities~\cite{7842423}, IoT services~\cite{8726071}.   & Wide coverage, low cost, flexible deployment.  &Energy constrained, unstable link, limited capacity. \\\hline
Terrestrial edge computing     &MEC~\cite{ETSIMEC}, Fog Computing~\cite{8100873}.        & N.A.                          &Limited.  &1-5ms (5G).                              &AR/VR gaming.  & High data rate, rich resources.  & Vulnerable to natural disasters, limited coverage.\\\hline
\end{tabular}
\end{table*}
\end{footnotesize}

\section{State-of-the-art orbital edge computing and aerial edge computing technologies}{\label{sec:review}}
In this section, we will review state-of-the-art orbital edge computing and aerial edge computing techniques.
TABLE~\ref{tab:Comparison} summaries some available edge computing technologies for SAGINs, and compares their advantages and challenges.

\subsection{Orbital Edge Computing}
Thanks to the global coverage of orbital network, implement edge computing in orbital network can improve the QoS of end users.

\subsubsection{Architecture}{\label{sec:space_architecture}}
Most of existing works rely on a ground control architecture: terrestrial stations send task request to the space, and satellites reply raw data (e.g., weather and radar maps) back to the ground for further analysis.
However, the satellites' communication links can be overloaded as the amount of raw data increases.
Thus, it is beneficial to bring edge computing functions in orbit to handle the processing capabilities of the raw data right near where it's collected, rather than sending it to the ground.
This type of data processing in orbit is known as orbital edge computing (or satellite/space edge computing), and has attracted great interest in recent years.
%A trend toward massive constellations of low Earth orbit (LEO) nanosatellites demands a new architecture for space systems.
%OEC
On the one hand, for the computation requirements in the space, authors in~\cite{OEC} propose an orbital edge computing (OEC) architecture which consists of earth-observing, camera equipped satellites and ground stations.
OEC is an edge-sensing and edge-computing system, provides edge computing to the satellites so that sensed raw data can be analyzed in orbit.
Thus, the transmission burden of downlinks can be alleviated.
Moreover, OEC also organizes the satellites into computational pipelines to reduce the edge processing latencies.
%Smec
On the other hand, for the computation requirements in the ground, authors in~\cite{8610431} propose an architecture of satellite mobile edge compuing (SMEC).
In the SMEC, end users in the ground without terrestrial edge servers can also enjoy satellite edge computing services via satellite links.
SMEC also offers computation offloading and content caching services, in order to alleviate the traffic burden of satellite-terrestrial network
(STN).
Thus, an end user in SMEC has the following computation offloading possibilities: proximal terrestrial offloading (PTO), satellite-borne offloading (SBO) and remote terrestrial offloading (RTO).

\subsubsection{Use Cases}{\label{sec:space_use_case}}
The orbital edge computing technique can be applied to various practical fields, such as Earth observation, space explorations, smart cities and emergency relief.
Take a real world application, Earth observation as an example.
Traditionally, radar and visual imagery gleaned from the space are transmitted in raw form to a terrestrial station, consume a lot of time and bandwidth.
On-orbit image recognition can reduce bandwidth and offer a processed image directly to the task requester.
The real-time imagery processing is critical for some applications, because the processed image may quickly loses its value if it cannot be analyzed in real-time, such as infrastructure monitoring, capturing debris and disaster relief.
To this end, researchers at Tokyo Tech have designed an edge-computing-based star tracker and an Earth sensor~\cite{Tokyo_Tech}.
The star tracker is developed to use sensor-equipped nanosatellite for long-term Earth monitoring in orbit, and the Earth sensor performs real-time image recognition using deep learning that identifies land use and vegetation distribution.

Another potential application could be intelligent transportation systems.
Current intelligent transportation systems (e.g., vehicular networks) are mainly based on terrestrial infrastructure (e.g., 4G/5G/WiFi).
Although the terrestrial infrastructure can provide high-speed data rate, it faces the challenges of coverage, and it is vulnerable to natural disasters (e.g., earthquakes).
For example, the 4G cellular network has poor coverage in isolated mountainous areas, and easy to break down in disaster areas.
To this end, authors in~\cite{7981533} propose a software defined SAGINs framework to support wide-area, seamless and efficient vehicular services.
In their architecture, terrestrial networks in urban areas provide high-speed wireless connections to vehicular users, orbital networks support wide-area and seamless connectivity to space-assisted connected vehicles in rural areas, while aerial network can enhance the capacity in hot-spot areas.
Through this hierarchical network architecture, vehicular users can enjoy various vehicular services in a real-time and cost-effective manner.

Other potential use cases could include any requiring global coverage, and any that could benefit from edge computing in orbit, such as weather reporting and military missions.

\subsubsection{Challenges}{\label{sec:space_challenges}}
Although orbital edge computing has the advantage of global coverage, it still faces the following challenges.
\textbf{i) Physical damage,} radiation is the biggest threat to orbital edge computing in orbit, because RAM, CPUs and GPUs are vulnerable to solar flares and cosmic radiation.
Radiation can cause ``bit flips'' in RAM and reboot an application execution process.
Furthermore, the latest generation of chips pack more transistors onto a piece of silicon.
Note that chips are more vulnerable to radiation as fabrication processes get smaller.
Thus, the hardware of nanosatellite should be carefully designed.
\textbf{ii) Long propagation latencies,} latency and bandwidth are the main challenges for orbital edge computing, especially for GEO and MEO satellites based edge computing scenario.
Large one way delay usually degrades the performance of real-time applications.
For example, the one way delay for GEO is about 270 ms, and MEO has the one way delay of 110 ms~\cite{8368236}.
On the other hand, downlinks between satellites and terrestrial stations are usually unreliable.
The high mobility of satellites with respect to the terrestrial stations complicates orbital edge computing, and imposes limitations on link availability.
Thus, the intermittently available downlinks incur high latency in orbital edge computing.
\textbf{iii) Limited energy supply,}
for orbital edge computing, the energy supply of communication and data processing must be harvested from the space.
However, due to the small size and limits solar panel surface area of a nanosatellite, the energy supply and computing capacity of a nanosatellite are quite limited.
Thus, running high energy consumption applications (e.g., deep learning algorithm) in a single nanosatellite may not be practical.

\subsection{Aerial Edge Computing}{\label{sec:aerial_edge_computing}}
Recently, aerial edge computing which uses HAPs (e.g., unmanned aerial vehicles and balloons) as carriers for data transmission and task execution has attracted much attention.
Compared with terrestrial edge computing, aerial edge computing has the features of flexible deployment, agile management, large coverage and low cost.

\subsubsection{Architecture}{\label{sec:aerial_architecture}}
Most of existing works rely on an air-ground integrated architecture.
For example, authors in~\cite{8436041} propose a framework of air-ground integrated mobile edge network (AGMEN), where multiple UAVs are flexibly deployed and scheduled to provide wireless coverage, computation and storage capacities for edge computing.
In AGMEN, the UAVs can also cache popular contents/services in order to reduce the transmission burden.
The performance of AGMEN can be improved by jointly optimizing UAV scheduling and air-ground cooperation.
Authors in~\cite{8675170} propose a hybrid computing model UAV-Edge-Cloud, which integrates UAV swarm into edge/cloud computing to guarantee the high QoS of computation-intensive applications.
The UAV swarm can extend the capacity of UAV-Edge-Cloud and assist the communication and computing of the edge and cloud networks.
The authors also discuss some real world applications for UAV-Edge-Cloud, such as smart cities.

\subsubsection{Use Cases}{\label{sec:aerial_use_case}}
For the real world applications of aerial edge computing, most of existing research focuses on the UAV-assisted edge computing for Internet of Things (IoT) services, such as VR/AR gaming~\cite{8726071} and crowd sensing~\cite{7842423}.
By equipping varied sensors and computation units, UAVs can perform real time application execution, or serve as a fog computing platform to support IoT services.
For example, authors in~\cite{8726071} propose a novel aerial edge computing architecture, UAV-clustering-based multi-modal multi-task offloading (UAV-M3T) to support vehicular VR/AR gaming.
By leveraging AI-based decision making, the UAV-M3T architecture jointly optimizes the communication, computation and storage resource of the network, as well as the trajectory of UAV swarm.
Thus, the UAV swarm performs virtual scene processing and provide real-time high resolution videos for high speed vehicular users.

Other potential applications could be crowd sensing applications in smart cities, such as urban target tracking and monitoring urban traffic congestion.
For the crowd sensing tasks in smart cities, it is more efficient to process part of raw data on UAV swarms locally before transmitting to remote cloud server.
Authors in~\cite{7842423} propose a UAV-based IoT platform to support face recognition based crowd surveillance in smart cities.
In the platform, the sensed raw data gleaned from UAVs can be either offloaded to a MEC node, or processed locally onboard UAVs.
They also develop a testbed in LTE network to show their platform can quickly detecting and recognizing a suspicious person in a crowd.
The use case can also be used in the scenarios of collisions and obstacle avoidance in smart cities.

\subsubsection{Challenges}{\label{sec:aerial_challenges}}
Although aerial edge computing has the features of low cost and easy deployment, there are many challenges to resolve before effective use of HAPs can be made to provide edge computing services.
\textbf{i) Energy constrained,} small UAVs are powered by batteries, thus energy constrained.
On the other hand, larger batteries affect the weight and fly time of UAVs.
\textbf{ii) Unstable link,} UAVs are partially controlled by terrestrial controller on demand, and move with varying speeds.
This would cause intermittent links between UAVs and terrestrial controller.
Limited fly time of UAVs and dynamicity of the network may cause frequent handovers in a task execution process.
\textbf{iii) Limited capacity,} due to limited load capacity, UAVs can only carry limited edge computing device (e.g., CPU, GPU and storage) and batteries.
Thus, the computation, communication and storage capacities for each single UAV are quite limited.

\section{Edge Computing-enabled Space-Air-Ground Integrated Networks}{\label{sec:EC-SAGIN}}
%¼ò½é
In this section, we first introduce the architecture of the proposed edge computing-enabled space-air-ground integrated networks (EC-SAGINs).

\begin{figure}[t!]
    \centering
    \includegraphics[width=3.5in]{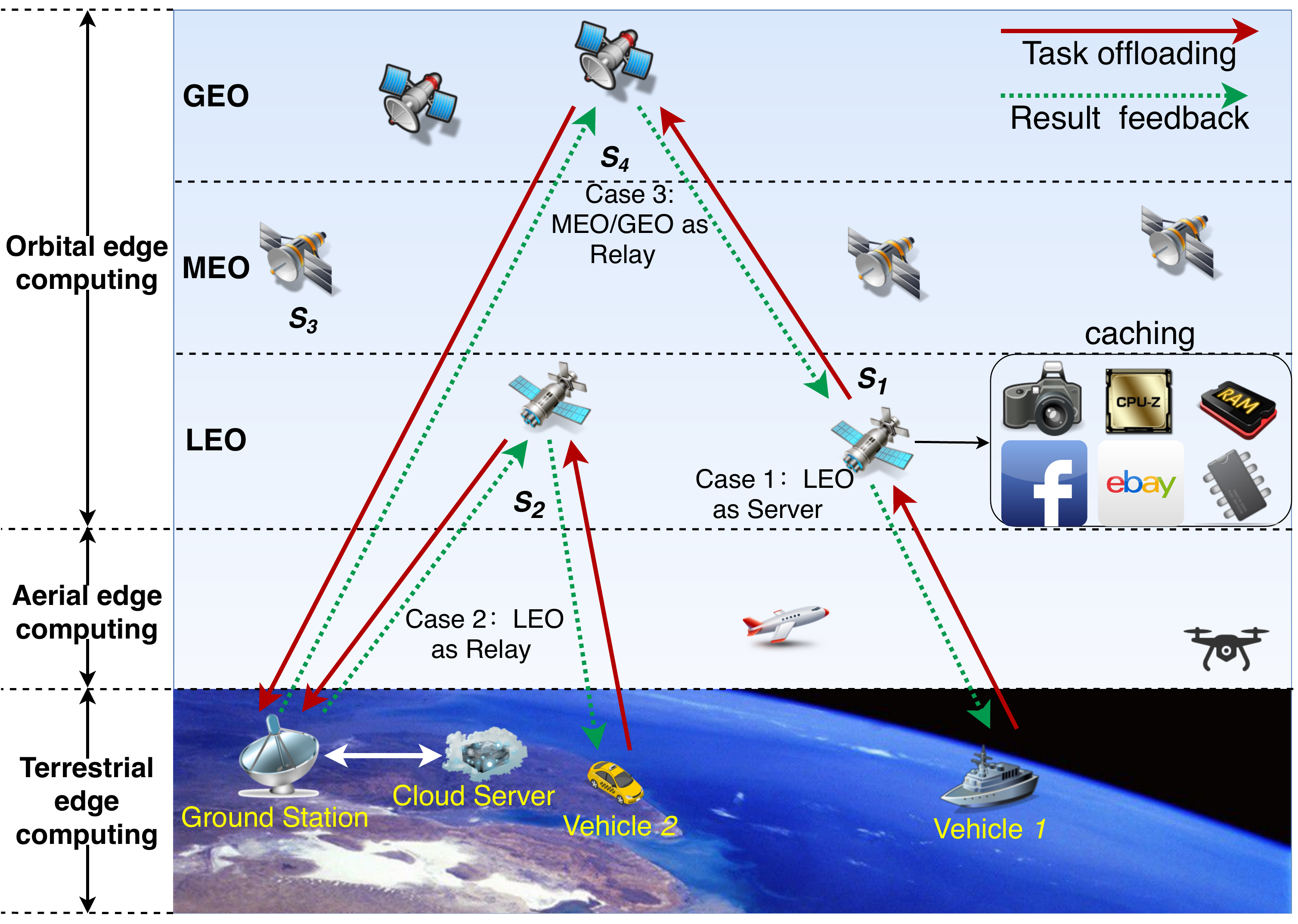}
    \caption{Architecture of Edge Computing-enabled Space-Air-Ground Integrated Networks (EC-SAGINs).}
    \label{fig:EC-SAGIN}
\end{figure}

\subsection{Architecture}{\label{sec:EC-SAGIN_architecture}}
EC-SAGINs are hierarchical edge computing structures consist of the following layers, as shown in Fig.~\ref{fig:EC-SAGIN}.

\subsubsection{LEO Layer}
The LEO layer of EC-SAGINs consists of LEO satellites are equipped with computation and storage resources (e.g., CPU and GPU) to support edge computing in orbit, so that raw data can be processed locally on satellites when the backhaul from satellites to the ground station are not available (i.e., LEO as server in Fig.~\ref{fig:EC-SAGIN}).
In addition, a LEO satellite can also act as a relay to forward the raw data to the ground station, if the satellite is in the coverage of the ground station, and resources of the satellite are occupied (i.e., LEO as relay in Fig.~\ref{fig:EC-SAGIN}).
Note that LEO satellites move quickly around the earth, thus the fronthaul between vehicles and the satellites are not always available.
The satellite coverage time (as will be explained in Section~\ref{sec:Coverage_Time}) must be considered when vehicles offload their tasks to LEO satellites.
At the same time, the computation, communication and storage resources for each single satellite are quite limited.
Thus, resource-efficient offloading strategies with fronthaul state consideration are required to reduce the task completion time for the LEO satellites.

\subsubsection{GEO and MEO Layers}
In EC-SAGINs, the GEO and MEO satellites play a key role in relaying raw data for either LEO satellites or terrestrial vehicles.
Note that in LEO network, backhaul limits satellites' datarate and connections, each LEO nanosatellite has to transmit raw data to a ground station by leveraging GEO/MEO.
For example, for the vehicle 1 in remote areas, as illustrated in Fig.~\ref{fig:EC-SAGIN}.
It is impossible for the vehicle to enjoy TEC services.
Thus, it can join EC-SAGINs when an edge computing enabled LEO satellite $S_{1}$ is close to him.
However, the LEO satellite $S_{1}$ is outside the coverage of the ground station.
Thus, the backhaul link between $S_{1}$ and the ground station will be unavailable.
%Thus, $NS_{1}$ cannot deliver data to target ground station until it reaches to location $L_{2}$.
By leveraging GEO relay, $S_{1}$ can first send data to GEO satellite $S_{4}$, then $S_{4}$ transmits the received data to the target ground station, as $S_{4}$ has a larger coverage and a closer distance to the ground station (i.e., MEO/GEO as relay in Fig.~\ref{fig:EC-SAGIN}).
Note that, relaying by GEO or MEO satellites can result in much higher transmission delay compared to LEO transmission.

\subsubsection{Aerial Layer}
The aerial layer consists of edge computing enabled high altitude platforms can offer broadband wireless connections and real-time edge computing services (e.g., fog computing) in the sky.
Note that HAPs support computation offloading and caching services to reduce the delay and energy consumption for vehicles.
However, the coverage and energy supply of the platforms are quite limited.
\subsubsection{Terrestrial Layer}
The terrestrial layer of EC-SAGINs consists of terrestrial edge computing facilities (e.g., MEC) and satellites' ground station.
The ground station can analyse the raw data received from LEO satellites, or send the data to a remote data center.

\subsection{Dynamic Offloading and Caching for EC-SAGINs}{\label{sec:EC-SAGIN_offloading}}
%¼ÆËãжÔØ
Based on the fact that terrestrial and aerial edge computing are relatively mature, thus we focus on the orbital edge computing of SAGINs in this article.
We propose a fine-grained joint offloading and caching scheme that is based on an orbit-ground collaboration.
Our objective is to provide real-time EC-SAGINs services for the terrestrial vehicles in the remote areas where TEC infrastructure is unavailable (e.g., rural areas and isolated islands).
Specifically, the vehicles first offload their tasks (either data transmission tasks or computational tasks) to their nearby LEO satellites.
Then, the satellites dynamically decide to offload the received data or not, according to the task state, network state and current available resources of the satellite (as will be explained in Section~\ref{sec:Problem-Formulation}).
At last, the satellites transmit the required data (either computation results or downloaded data) to the vehicles, and decide if to cache the required data for future reuse or retransmission.

\section{System Model}{\label{sec:System-Model}}
In this section, we will introduce the task model, satellite coverage time model, communication model, computing and caching model for the EC-SAGINs.
For ease of reference, key notations of our system model are listed in TABLE.~\ref{table:keynotations}.

\begin{table}[t]
\caption{Key Notation}
\label{table:keynotations}
 \begin{tabular}{m{1in}<{}m{2.1in}<{}}\hline
 \toprule
\textbf{Symbol}&\textbf{Definition}\\
  \midrule
$\mathcal{M}=\{1,2,...,M\}$           &The set of edge computing enhanced LEO satellites.\\
$o=(\mathcal{V},\mathcal{D})$         & Task model, where $\mathcal{V}$ represents the set of sub-tasks, and $\mathcal{D}$ denotes the data dependencies between the sub-tasks. \\
$\zeta_{v}$                           & The workload (CPU cycles/number of instructions to be executed) of sub-task $v$.\\
$d_{v}^{i}, d_{v}^{o}$                & Input and output of sub-task $v$, respectively.\\
$t_{m}^{c}$                           & The coverage time of satellite $m$. \\
$r_{vs}^{fh}, r_{sg}^{bh}$            & Vehicle-to-satellite (fronthaul) and satellite-to-ground (backhaul) data rates, respectively.\\
$p_{v}$, $p_{s}$                      & Transmission power of the vehicle and LEO satellites, respectively..\\
$d_{vs}$, $d_{sg}$                    & The propagation delay of the fronthaul and the relay delay of the backhaul, respectively.\\
$\boldsymbol K$                       & A caching placement strategy.\\
$\mathcal{S}(t)$                      & System state space at time slot $t$.\\
$\mathcal A(t)$                    & System action space at time slot $t$.\\
$\boldsymbol a^{of}, \boldsymbol a^{ch}$ & Fine-grained offloading decision and caching placement strategies.\\
  \bottomrule
 \end{tabular}
\end{table}

Assume that the network model of EC-SAGINs consists of: i) a vehicle in remote areas (e.g., desert, rural areas, isolated islands and deep-sea), ii) a set $\mathcal{M}=\{1,2,...,m,...,M\}$ of edge computing enabled LEO satellites, iii) MEO/GEO relay satellites and iv) a ground cloud server, as shown in Fig.~\ref{fig:EC-SAGIN}.
Note that the vehicle in this article includes i) terrestrial vehicles, such as automobile and truck, ii) aerial vehicles, such as UAVs, and iii) marine vehicles, such as liner and warship.
From the perspective of the vehicles in remote areas, it is impossible for them to enjoy cloud servers through terrestrial edge computing (TEC) infrastructures (e.g., 5G base stations and roadside units).
For example, the 4G networks have poor coverage in rural areas, and most isolated islands in the world have no TEC infrastructures~\cite{7981533}.
Moreover, TEC infrastructures are usually vulnerable to natural disasters (e.g., earthquakes and flood).
Thus, rescue vehicles in disaster-stricken areas may lose Internet connections.
To this end, the vehicles in remote areas prefer to join the EC-SAGINs, in order to enjoy Internet access and cloud services.

\subsection{Task Model}

\begin{figure}[t!]
    \centering
    \includegraphics[width=3in]{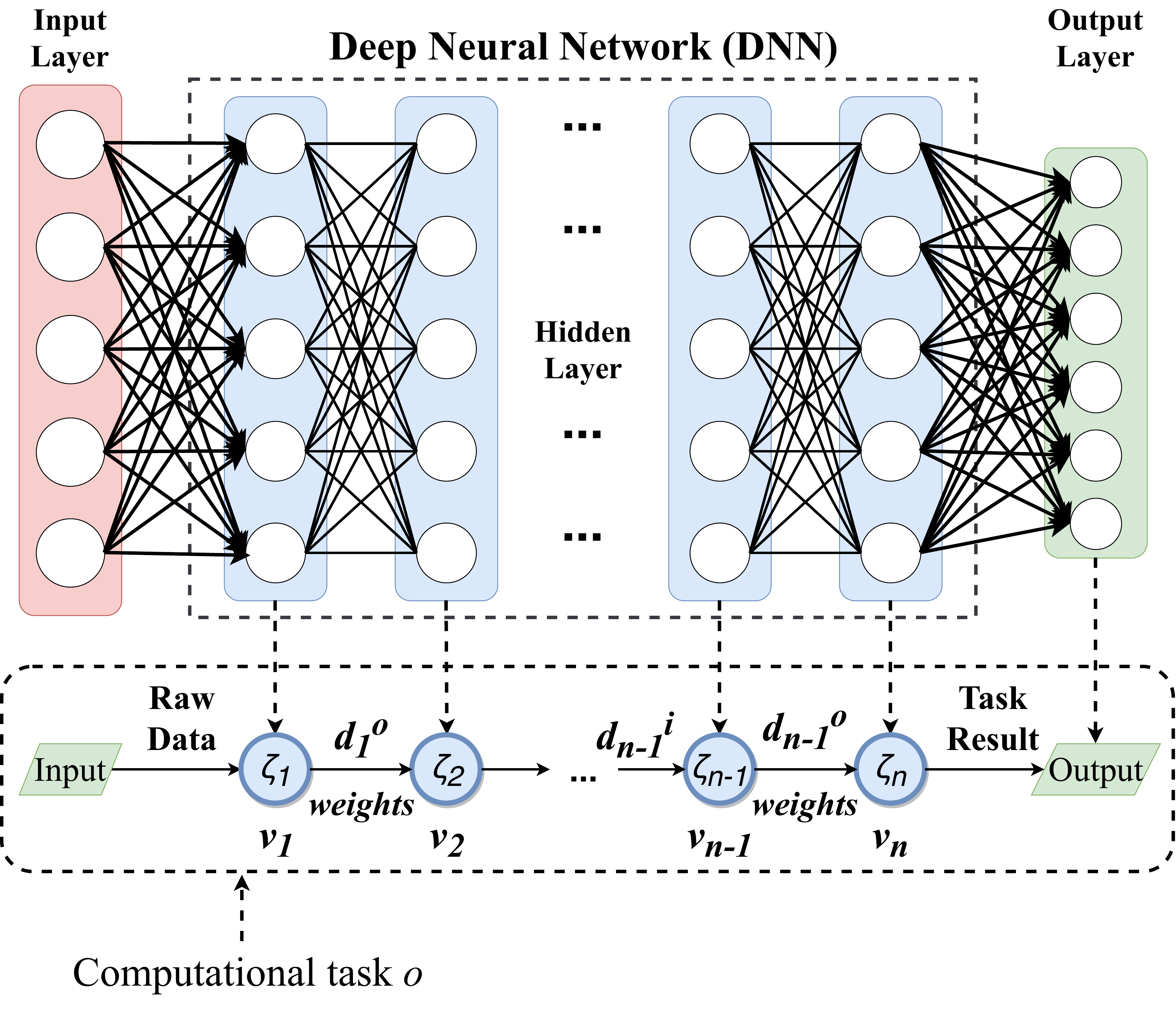}
    \caption{An illustrative example of DNN model partition.}
    \label{fig:DNN}
\end{figure}

In this paper, we consider a fine-grained task partitioning manner~\cite{MAUI}, and split a computational task $o, (o\in\mathcal{O})$ into multiple sub-tasks.
Let $o=(\mathcal{V},\mathcal{D})$, where $\mathcal{V}$ represents the set of sub-tasks, and $\mathcal{D}$ denotes the data dependencies between the sub-tasks.
Let $|\mathcal{V}|$ represent the number of sub-tasks in task $o$.
After receiving the task requirement of vehicle, LEO satellite $m$ can offload sub-task $v$ $(v\in \mathcal{V})$ to the ground station, or executed locally in the satellite.
Take the deep neural network (DNN) model partition~\cite{8736011} as an example.
In order to reduce the burden of edge intelligence applications execution on end devices, DNN model partition can split a DNN training phase (i.e., the computational task $o$) into multiple sub-tasks (i.e., the hidden layers of DNN), as shown in Fig.~\ref{fig:DNN}.
By dynamically offloading the resource intensive sub-tasks to a more resourceful server (i.e., the ground station in EC-SAGINs), DNN model partition can speed up the inference process.

Moreover, we consider a more general scenario that each sub-task $v$ ($v\in\mathcal{V}$) can be either a computational task or a data transmission (uploading/downloading) task.
Let a parameter tuple $k_{v}=\left \langle \zeta_{v}, d_{v}^{i}, d_{v}^{o}\right \rangle$ characterize the sub-task $v$, where $\zeta_{v}$ ($v\in\mathcal{V}$) refers to the workload of sub-task $v$.
$d_{v}^{i}$ represents the input data of sub-task $v$, and $d_{v}^{o}$ refers to the output.

$\rho_{v}$ indicates the required CPU cycles a CPU core will perform per byte for the input data processed by the sub-task $v$ (i.e., complexity of task $o$).
Thus, we have $\zeta_{v}=\rho_{v} \cdot d_{v}^{i}$.
Note that $\zeta_{v}$ is decided by the algorithm complexity of sub-task $v$.

\subsection{Satellite Coverage Time Model}{\label{sec:Coverage_Time}}
Compared with MEO and GEO satellites, LEO satellites have lower orbital altitudes but higher velocity.
According to the works in~\cite{8675467,6174409}, LEO satellites only enable to server vehicles in a certain period of time (i.e., satellite coverage time).
Form the perspective of the vehicle, the satellite coverage time of LEO satellite $m$ ($m\in\mathcal{M}$) denoted by $t_{m}^{c}$ is given as follows:
\begin{equation}
\begin{aligned}
\label{eq:satellite_coverage_time}
t_{m}^{c}=&\frac{1}{\eta}arccos\left(\frac{cos\theta_{0}}{cos\theta_{m}}\right),\\
\eta=&\frac{\eta_{m}-\eta_{e}cosi}{2}.
\end{aligned}
\end{equation}
Note that $\eta$ is an attribute of the LEO satellite, where $\eta_{m}$ denotes the angular velocity of satellite $m$, $\eta_{e}$ is the angular velocity of the earth's rotation, $i$ represents the inclination of circular orbits.
$\theta_{0}$ in (\ref{eq:satellite_coverage_time}) refers to the earth central angle with respect to the minimum elevation angle, and can be calculated as follows:
\begin{equation}
\theta_{0}=arccos\left(\frac{r_{e}}{r_{e}+h}cos\xi_{0}\right)-\xi_{0},
\end{equation}
where $h$ is the satellite altitude, $r_{e}$ refers to the radius of the earth.
$\xi_{0}$ denotes the minimum elevation angle required for visibility.
$\theta_{m}$ in (\ref{eq:satellite_coverage_time}) denotes the the minimal value of angular distance between the LEO satellite $m$'s ground trace and the vehicle's location.

From the perspective of the vehicle, its remaining satellite coverage time $t_{m}^{c}$ in the pass of LEO satellite $m$ is decided by the value of $\theta_{m}$.

\begin{table*}[!t]
\caption{Pre-classified offloading and caching actions}
\label{tab:Pre-classified}
\centering
\renewcommand\arraystretch{1.2}
\begin{tabular}{|m{20mm}<{\centering}|m{5mm}<{\centering}|m{5mm}<{\centering}|m{5mm}<{\centering}|m{30mm}<{\centering}|m{30mm}<{\centering}|m{20mm}<{\centering}|m{20mm}<{\centering}|}\hline
Task category                    & $\zeta_{s}$    & $d_{v}^{i}$         &  $d_{v}^{o}$     & $T_{o}^{cpl}$ (cache not hit)             & $T_{o}^{cpl}$ (cache hit)  &Available $\{a_{v}^{of},a_{v}^{ch}\}$ $(T_{sv}(d_{v}^{o})+d_{vs}<t_{m}^{c})$                   &Available $\{a_{v}^{of},a_{v}^{ch}\}$ $(T_{sv}(d_{v}^{o})+d_{vs}>t_{m}^{c})$\\\hline
Data transmission (uploading)    & $=0$           & $>0$                &  $=0$            & $\sum_{v\in\mathcal{V}}(\frac{d_{v}^{i}}{r_{vs}^{fh}}+d_{vs}+\frac{d_{v}^{i}}{r_{sg}^{bh}}+d_{sg})$           & $\sum_{v\in\mathcal{V}}(\frac{d_{v}^{i}}{r_{vs}^{fh}}+d_{vs}+\frac{d_{v}^{i}}{r_{sg}^{bh}}+d_{sg})$                & 10,11                    &  10,11 \\\hline
Data transmission (downloading)  & $=0$           & $=0$                &  $>0$            & $\sum_{v\in\mathcal{V}}(\frac{d_{v}^{o}}{r_{sg}^{bh}}+d_{sg}+T_{sv}(d_{v}^{o})+d_{vs})$    & $\sum_{v\in\mathcal{V}}(T_{sv}(d_{v}^{o})+d_{vs})$                   & 00,01                    &  01 \\\hline
Computation                      & $>0$           & $>0$                &  $>0$            & $\sum_{v\in\mathcal{V}}\{\frac{d_{v}^{i}}{r_{vs}^{fh}}+d_{vs}+a_{v}^{of}\cdot(D_{m,v}^{off}+d_{sg})+(1-a_{v}^{of})\cdot D_{m,v}^{loc}+T_{sv}(d_{v}^{o})+d_{vs}\}$                 & $\sum_{v\in\mathcal{V}}(\frac{d_{v}^{i}}{r_{vs}^{fh}}+d_{vs}+T_{sv}(d_{v}^{o})+d_{vs})$                   & 00,01,10,11                    &  01,11 \\\hline
\end{tabular}
\end{table*}

\subsection{Communication Model}
In EC-SAGINs, we consider two communication interfaces, i.e., vehicle-to-satellite and satellite-to-ground, respectively.
Note that the above interfaces use different spectrum bands, thus lead to no interference to each other~\cite{345873}.
Then, the delay performance of data transmission is analyzed in detail.

\subsubsection{Vehicle-to-satellite}
In EC-SAGINs, the vehicle communications with a LEO satellite through Ku-band~\cite{8368236}, since the frequency band has no interference to terrestrial wireless communication systems (e.g., 4G, 5G and Wifi).
However, the channel condition of Ku-band is easily affected by the communication distance and the rain attenuation~\cite{7821087}.
Therefore, the data rate of the vehicle-to-satellite link (i.e., fronthaul link from the vehicle to the LEO satellite) denoted by $r_{vs}^{fh}$ is given by:
\begin{equation}
r_{vs}^{fh}=\Lambda B_{vs}\log_{2}\left(1+\frac{p_{v}\cdot|h_{vs}|^{2}}{\sigma^{2}}\right),
\end{equation}
where $B_{vs}$ denotes the channel bandwidth of the vehicle-to-satellite link, $p_{v}$ is the transmission power of the vehicle, $\sigma^{2}$ indicates the power of noise, $h_{vs}$ denotes the channel fading coefficient between vehicle and LEO satellite $m$.
$\Lambda$ refers to the rain attenuation ratio.
The estimation of this value has been studied in~\cite{633864,8408531} which is thus beyond the scope of our work.
For simplicity, we set the rain attenuation ratio to a fixed value in this article.

Note that the transmission delay is greatly affected by the communication distance between the LEO satellite and the vehicle.
Thus, the propagation delay cannot be ignored and be defined as $d_{vs}$.

\subsubsection{Satellite-to-ground}
In EC-SAGINs, LEO satellites offload data or computational tasks directly to the ground station or through MEO/GEO relay satellites.
The satellite-to-ground backhaul link occupies a wider bandwidth than the fronthaul link by leveraging Ka-band~\cite{6735634}.
Ka-band can also be used in MEO/GEO satellite systems to provide highspeed data transmission (e.g., video and movie).
Therefore, the data rate of the satellite-to-ground link denoted by $r_{vs}^{fh}$ is given by:
\begin{equation}
r_{sg}^{bh}=\Lambda B_{sg}\log_{2}\left(1+\frac{p_{s}\cdot|h_{sg}|^{2}}{\sigma^{2}}\right),
\end{equation}
where $B_{sg}$ denotes the channel bandwidth of the satellite-to-ground link, $p_{s}$ is the transmission power of the LEO satellite $m$, $\sigma^{2}$ indicates the power of noise, $h_{sg}$ denotes the channel fading coefficient between LEO satellite $m$ and the ground station.

Note that the LEO satellite $m$ has to offload tasks through MEO/GEO relay satellites, if $m$ is outside the coverage of the ground station.
Thus, relay delay cannot be ignored and be defined as $d_{sg}$.

\subsection{Computing and Caching Model}
Under the EC-SAGINs framework, the vehicle first offloads a vehicular task $o$ to LEO satellite $m$.
Then, the satellite $m$ has to i) upload or download the related data $d_{v}^{i}$ to the ground station, if sub-task $v$ is a data transmission task, or ii) process the data locally or offload the data to the ground station, if sub-task $v$ is a computation task.
Let $|d_{v}^{i}|$ denote the size (in bytes) of $d_{v}^{i}$.
Thus, the delay for satellite $m$ to transmit the related data $d_{v}^{i}$ to the ground station is given as $D_{m,v}^{off}=\frac{|d_{v}^{i}|}{r_{sg}^{bh}}$.
Similarly, the delay for satellite $m$ to process the related data $d_{v}^{i}$ locally is given as $D_{m,v}^{loc}=\frac{\rho_{v} \cdot |d_{v}^{i}|}{f_{m}}$, where $f_{m}$ represents the CPU frequency of LEO satellite $m$.

After receiving the task output $d_{v}^{o}$ (i.e., computation results or downloaded data), the LEO satellite has to decide if cache the data locally for future reuse or retransmission.
Assume that a library consists of $L$ task outputs, given as $\mathcal{L}=\{d_{v}^{o}, v\in\mathcal{V}, o\in\mathcal{O}\}$.
The popularity distribution of the outputs conditioned on the event that the vehicle makes a request (i.e., reuse or retransmission) for its current sub-task $v$.
%We also assume that LEO satellite $m$ allocates a finite-capacity storage $C$ for the vehicle to cache the results.

Then, a popularity-based caching policy~\cite{Zipf} is used to evaluate the request probability for a task output $d_{v}^{o}$ as follows:
\begin{equation}{\label{eq:Zipf}}
\begin{aligned}
f(d_{v}^{o},\delta,L)=\frac{1}{(d_{v}^{o})^{\delta}}\sum\limits_{l=1}^{L}\frac{1}{l^{\delta}},
\end{aligned}
\end{equation}
where $\delta$ indicates the skewness of the popularity profile.
Note that the popularity profile is uniform over files when $\delta=0$, and becomes more skewed with the growth of $\delta$~\cite{Zipf}.
Let a matrix $\boldsymbol K=[k_{1},k_{2},...,k_{L}]_{1\times L}$ denote a caching placement strategy where $k_{l}=1$ $(l=1,2,...,L)$ represents the related task output is cached in the satellite, and $k_{l}=0$ indicates the output is not cached.

\begin{figure*}[t!]
    \centering
    \includegraphics[width=6in]{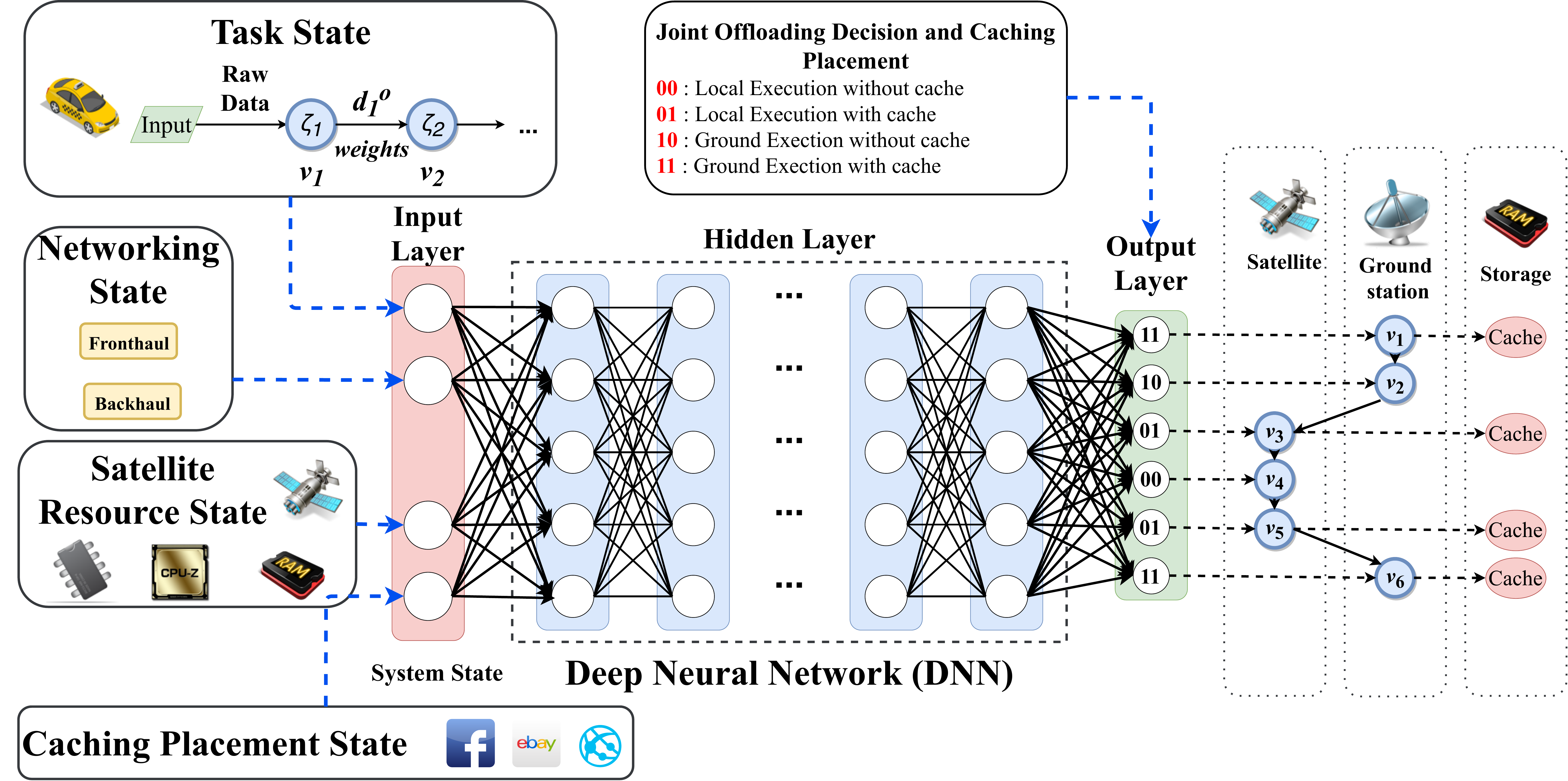}
    \caption{Proposed deep imitation learning-based decision making model.}
    \label{fig:deep_rl}
\end{figure*}
\section{Problem Formulation}{\label{sec:Problem-Formulation}}

In this section,  we describe the optimization problem for the EC-SAGINs framework.
The main objective of the framework is to minimize the task completion time as well as the satellite resource usage.
Therefore, we model the problem as reinforcement learning scenario or Markov Decision Processes (MDPs)~\cite{MDP2}, and introduce the space state, action state and reward function for the problem.

\subsection{State Space}{\label{sec:State_space}}
At time slot $t$, the state space includes task state, networking state, satellite state and caching placement state.
Thus, state space $\mathcal{S}(t)$ is represented as follows:
\begin{equation}{\label{eq:state_space}}
\mathcal{S}(t)=\{o,t_{m}^{c},r_{vs}^{fh},r_{sg}^{bh},d_{vs},d_{sg},f_{m},p_{v},p_{s},\boldsymbol K\}.
\end{equation}

\subsection{Action Space}{\label{sec:Action_space}}
Under the EC-SAGINs framework, the LEO satellite needs to make a fine-grained offloading and caching placement decision for the current task $o$.
Let a matrix $\boldsymbol A(t)=\{\boldsymbol a^{of}, \boldsymbol a^{ch}\}$ represent the decision ($\boldsymbol A(t)\in\mathcal{A}$).
$\mathcal{A}$ represents the system action space.
$\boldsymbol a^{of}=[a_{1}^{of},a_{2}^{of},...,a_{|\mathcal{V}|}^{of}]_{1\times|\mathcal{V}|}$ denotes a fine-grained offloading decision matrix for the task $o$, where $a_{v}^{of}=1$ $(v=1,2,...,|\mathcal{V}|)$ denotes to offload sub-task $v$ to the ground station, and $a_{v}^{of}=0$, not to offload.
$\boldsymbol a^{ch}=[a_{1}^{ch},a_{2}^{ch},...,a_{|\mathcal{V}|}^{ch}]_{1\times|\mathcal{V}|}$ denotes a fine-grained caching placement matrix for the task $o$, where $a_{v}^{ch}=1$ $(v=1,2,...,|\mathcal{V}|)$ denotes to cache the output $d_{v}^{o}$ of sub-task $v$, and $a_{v}^{ch}=0$, not to cache.

\subsection{State Transition Probabilities}
The probabilities associated with different system states changes are called state transition probabilities as follows:
\begin{equation}{\label{eq:Transition_Probabilities}}
\text{Pr}\left(\mathcal{S}(t+1)|\mathcal{S}(t),\boldsymbol A(t)\right).
\end{equation}
It represents the probability of state $\mathcal{S}(t+1)$ in the current time slot $t+1$, if action $\boldsymbol A(t)$ is taken at system state $\mathcal{S}(t)$ in the previous time slot.

\subsection{Reward Function}{\label{sec:Reward_Function}}
Note that the vehicle needs to pay the usage of wireless spectrums, computation and storage fees to the satellite.
Thus, the main objective of the EC-SAGINs framework is to minimize the task completion time as well as the satellite resource usage.
To optimize the performance of the EC-SAGINs framework, a proper offloading and caching decision $\boldsymbol A^{*}$ ($\boldsymbol A^{*}\in\mathcal{A}$) for the task $o$ must be made, in order to minimize the resource usage and task completion time as follows:
\begin{equation}
\begin{aligned}
\label{eq:Reward}
R(t)=&R^{comp}(t)+R^{comm}(t)+R^{cache}(t)+R^{cpl}(t)\\
=&\pi^{comp}\times\sum_{v\in\mathcal{V}}(1-a_{v}^{of})\cdot\zeta_{v}+\pi^{comm}\times\sum_{v\in\mathcal{V}}a_{v}^{of}\cdot d_{v}^{i}+\\
&\pi^{cache}\times\sum_{v\in\mathcal{V}}(a_{v}^{ch})\cdot d_{v}^{o}+\pi^{cpl}\times T_{o}^{cpl},
\end{aligned}
\end{equation}
where $R(t)$ denotes the current system reward, $R^{comp}(t)$, $R^{comm}(t)$, $R^{cache}(t)$ and $R^{cpl}(t)$ represent the rewards of computation resource usage, communication resource usage, storage resource usage and task completion time, respectively.
$\pi^{comp}$, $\pi^{comm}$, $\pi^{cache}$ are the price for the computation, communication and storage resources, respectively.
$T_{o}^{cpl}$ denotes the task completion time for task $o$, as shown in TABLE~\ref{tab:Pre-classified}.
$\pi^{cpl}$ refers to a weight factor.

\section{Algorithm Design}{\label{sec:Algorithm_Design}}
In this section, we will present the algorithm for the optimization problem.
The algorithm consists a pre-classified offloading and caching scheme and a deep imitation learning (DIL) based decision making scheme, respectively.

\subsection{Pre-classified Offloading and Caching Scheme}
According to the offloading and caching matrix $\boldsymbol A(t)$, we can find that the action space increases exponentially with the growth of the sub-task number $|\mathcal{V}|$.
Indeed, the complexity for achieving the optimal decision is $o(4^{|\mathcal{V}|})$.
Thus, it is hard to obtain the optimal solution in polynomial time.
To this end, we first pre-classify the actions according to the system state, in order to decrease the size of action space $|\mathcal{A}|$.
TABLE~\ref{tab:Pre-classified} illustrates the pre-classified offloading and caching scheme.
The scheme pre-classifies the offloading and caching actions of a sub-task $v$ (i.e., $\{a_{v}^{of},a_{v}^{ch}\}$) by removing impossible actions.
For example, for a data uploading sub-task $v$, all the input data $d_{v}^{i}$ must upload to the ground server.
Thus, the available offloading and caching actions should be ``10'' and ``11''.
Similarly, for a data downloading sub-task, since the size of input data equals to 0.
Therefore, the offloading action must set to $a_{v}^{of}$ be $0$.
In addition, the satellite must cache the output data for retransmission, if the time spent for the satellite transmit the output to the vehicle (i.e., $T_{sv}(d_{v}^{o})+d_{vs}$) is larger than the remaining satellite coverage time $t_{m}^{c}$.
In this scenario, the caching placement action $a_{v}^{ch}$ is set to be $1$, as shown in TABLE~\ref{tab:Pre-classified}.

Through the action pre-classification scheme, the action space of our optimization problem can be greatly reduced.
In the next step, we will propose a deep imitation learning (DIL) based decision making scheme, as shown in Fig.~\ref{fig:deep_rl}.

\subsection{Deep Imitation Learning-based Decision Making Scheme}

\begin{figure}[t!]
    \centering
    \includegraphics[width=3in]{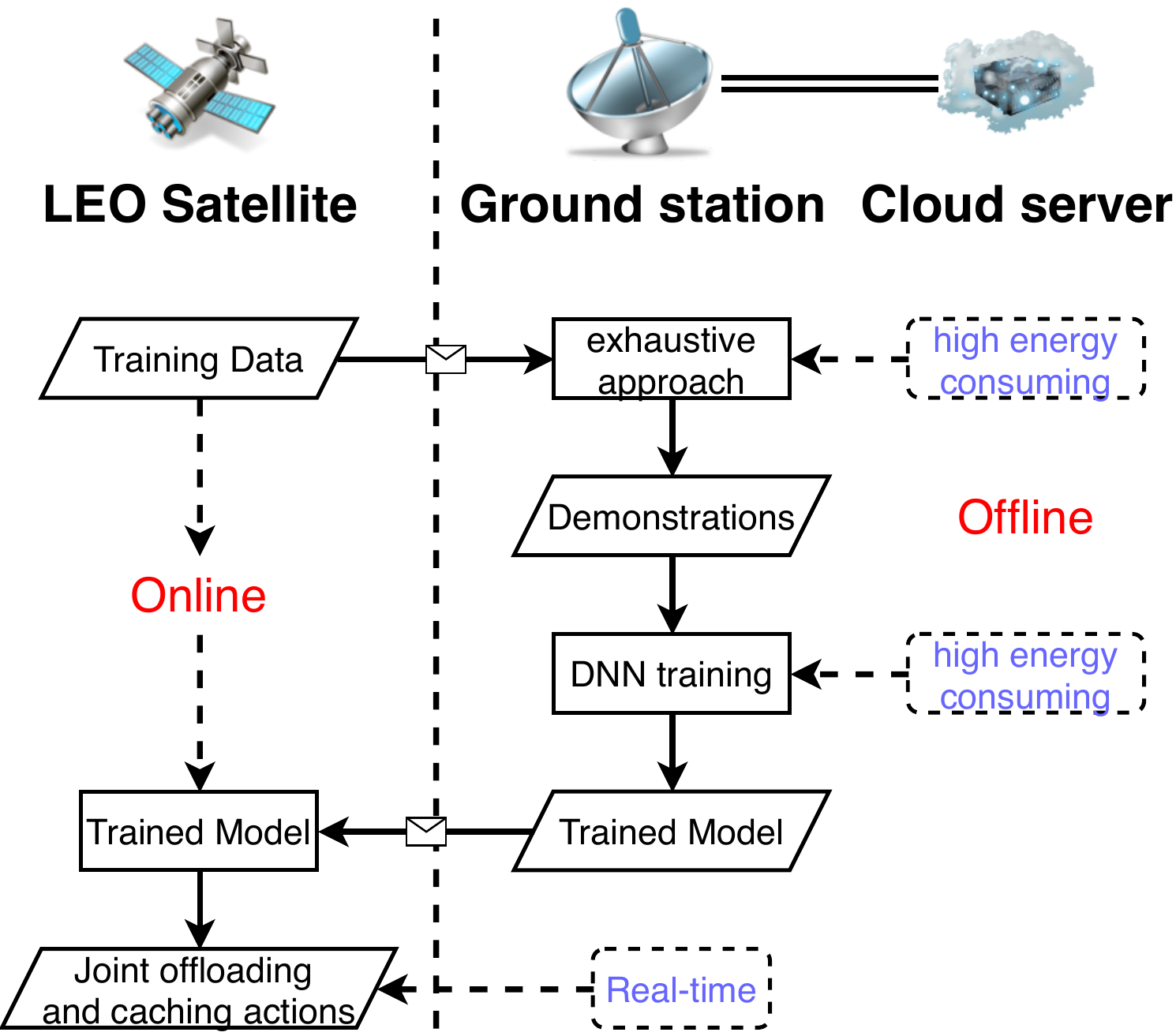}
    \caption{Workflow for the deep imitation learning-based decision making scheme.}
    \label{fig:workflow}
\end{figure}

Current studies on deep learning based intelligent offloading or caching schemes generally concentrate on the deep reinforcement learning (DRL) approaches~\cite{8941121,8434316,9026935,8879573}.
However, the online training manner of DRL is usually computation intensive and high energy consuming.
As stated earlier, due to the small size and limits solar panel surface area of LEO satellites, it is hard for the satellites to run deep learning algorithms.
To this end, we take a deep imitation learning based model training approach where the model is trained in the ground in an offline manner.
The workflow of the approach is illustrated in Fig.~\ref{fig:workflow}.

Through behavioral cloning, imitation learning aims to learn an optimal policy through imitating the demonstrations~\cite{DIL}.
Thus, we first generate optimized demonstrations to train our model.
Specifically, the satellite first sends the collected training data (i.e., system state) to the ground.
Then, we generate a lot of optimal decision samples $\boldsymbol A^{*}$ by solving (\ref{eq:Reward}).
Note that, generating the samples is time and energy consuming, thus the process is completed by the ground station in an offline manner, as shown in Fig.~\ref{fig:workflow}.

Then, deep neural network (DNN) is used to extract and train the features of the demonstrations in an offline manner.
As shown in Fig.~\ref{fig:deep_rl}, the inputs of our model is the system state, and the outputs is joint offloading and caching actions for a task.
Note that the output layer is composed of $|\mathcal{V}|$ neurons that indicates the joint offloading and caching actions of the $|\mathcal{V}|$ sub-tasks.
The traditional rectified linear unit (ReLU) is used as activation function for the hidden layers.
We also use the sigmoid function as the output of the model.
Cross-entropy loss~\cite{Nam2014} is considered to evaluate the performance of model.
We optimize the neural network by leveraging Adam optimizer~\cite{8624183}.
The computation intensive DNN training process is also done by the cloud server in the ground.
After the above offline model training, the ground station send the trained model to the LEO satellite.
Thus, the satellite can make real-time offloading and caching actions in an on-line and energy efficient manner.

Although satellite edge computing still suffers from large propagation delay, it can serve as an enabling technology in 6G networks.
On the one hand, satellite edge computing is regarded as a supplement to terrestrial edge computing in remote areas.
Note that, for most real-world applications, their delay requirements in remote areas are not as high as those in urban areas.
Take the application of autonomous vehicles as an example.
The distance between vehicles in remote areas are usually larger than that in urban areas.
Thus, vehicles in remote areas can tolerate a higher execution delay to avoid collision.
On the other hand, the large transmission delay of satellites can be alleviated by exploiting line-of-sight (LoS) connectivity and emerging optical communication technologies.
For example, an 8-beam free space optical link enables high-speed point-to-point connectivity with a peak rate of 1 Gb/s.
Low-cost white LEDs can also be utilized to achieve data rate over 500 Mb/s~\cite{8436052}.
With the development of satellite and communication technologies, the delay can be further reduced in the future.
\section{Performance Evaluation}{\label{sec:performance_evaluation}}
In this section, we evaluate the performance of our EC-SAGINs.
Specifically, we use MATLAB deep learning toolbox~\cite{matlab_rl} to implement the algorithm.
First, we describe the simulation environment and introduce the related benchmark strategies.
Then, we compare the performance of the proposed DIL-based algorithm with the benchmark strategies and discuss simulation results.
The values of the parameters are summarized in Table~\ref{tab:Parameters}.

\begin{footnotesize}
\begin{table}[tp]
\caption{Network Parameters}
\label{tab:Parameters}
\centering
\renewcommand\arraystretch{0.9}
\begin{tabular}{|m{17mm}<{\centering}|m{17mm}<{\centering}||m{17mm}<{\centering}|m{17mm}<{\centering}|}\hline
Parameter&Value&Parameter&Value\\\hline
$B_{vs}$         &2MHz               &$B_{sg}$                                  & 3MHz                   \\\hline
Learning rate    &0.001        &Discount factor                           & 0.95                 \\\hline
Optimizer        &Adam             &Activation function                   & ReLU                \\\hline
$L$              &30              &Hidden layer                           & 3                 \\\hline
$f_{m}$          &$10^{10}$ cycles/s               &$d_{vs}$                               & 30 ms                   \\\hline
$d_{sg}$         &270 ms                           &$\mathcal{D}$                          & [100, 500] KB                \\\hline
$\rho_{s}$       &[0, 12000] cpb                   &$\Lambda$                              & 0.8               \\\hline
$\delta$         &1                                &$t_{m}^{c}$                              & 5 min               \\\hline
\end{tabular}
\end{table}
\end{footnotesize}

Assume the EC-SAGINs framework consists of a vehicle, a LEO satellite, a GEO relay satellite and a ground station, as shown in Fig.~\ref{fig:EC-SAGIN}.
The computation capacity $f_{m}$ for the LEO satellite is $10^{10}$ cycles/s.
The propagation delay $d_{vs}$ between LEO satellite and ground station is 30 ms, and the one way delay $d_{sg}$ between GEO relay satellite and the ground station is 270 ms as shown in TABLE~\ref{tab:Comparison}.
For the task model, we assume that it is composed of 6 sub-tasks, and the data dependency $\mathcal{D}$ is based on a uniform distribution in the range of [100, 500] KB.
The value $\rho_{s}$ for sub-tsak $s$ follows the uniform distribution in the range of [0, 12000] cpb.

\begin{figure}[t!]
    \centering
    \includegraphics[width=3in]{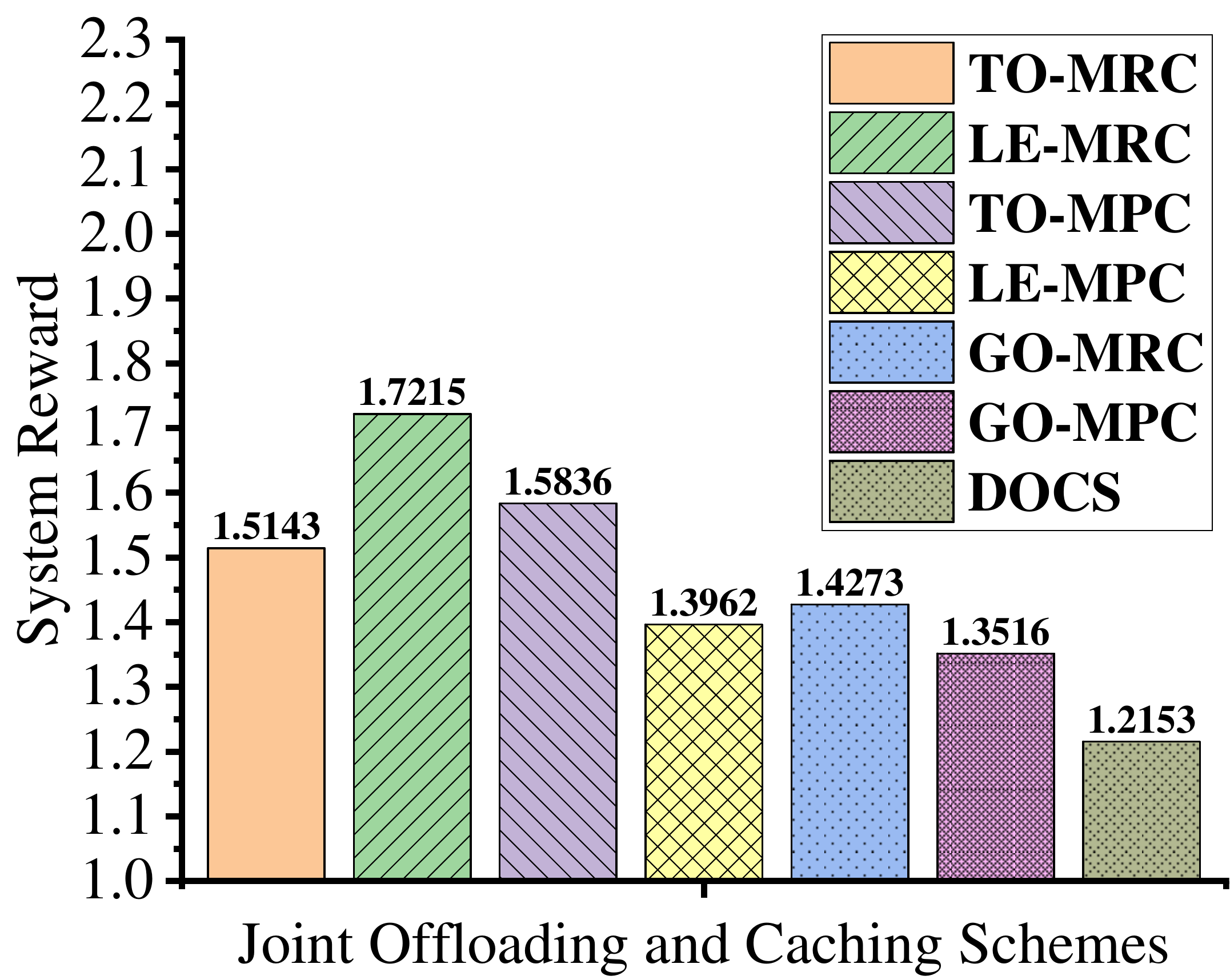}
    \caption{System reward performance for different offloading and caching schemes.}
    \label{fig:reward}
\end{figure}

Then, we implement our deep imitation learning-based offloading and caching scheme (DOCS), and compare it with the following benchmark offloading and caching schemes, namely:
\begin{itemize}
\item \emph{Local Execution scheme (LE):} A coarse-grained computation offloading scheme, which means that all the sub-tasks are processed on the satellite locally.
\item \emph{Total Offloading scheme (TO):} A coarse-grained computation offloading scheme, which means that all the sub-tasks are offloaded to the ground station.
\item \emph{Greedy-based Offloading scheme (GO):} LEO satellite chooses the sub-action that can maximize the offloading cost in each sub-task execution step.
\item \emph{Most Recent Caching scheme (MRC):} LEO satellite chooses to cache the most recent task outputs until reach to the storage capacity.
\item \emph{Most Popular Caching scheme (MPC):} LEO satellite chooses to cache the most popular  task outputs until reach to the storage capacity.
\end{itemize}

Fig.~\ref{fig:reward} reports the system reward performance for different offloading and caching schemes.
Note that our DOCS outperforms other benchmark offloading and caching schemes.
In fact, our DOCS can reduce the system reward on average by $19.74\%$, $29.40\%$, $23.26\%$, $12.95\%$, $13.85\%$ and $10.08\%$ compared to the TO-MRC, LE-MRC, TO-MPC, LE-MPC, GO-MRC and GO-MPC schemes, respectively.

\begin{figure}[t!]
    \centering
    \includegraphics[width=3in]{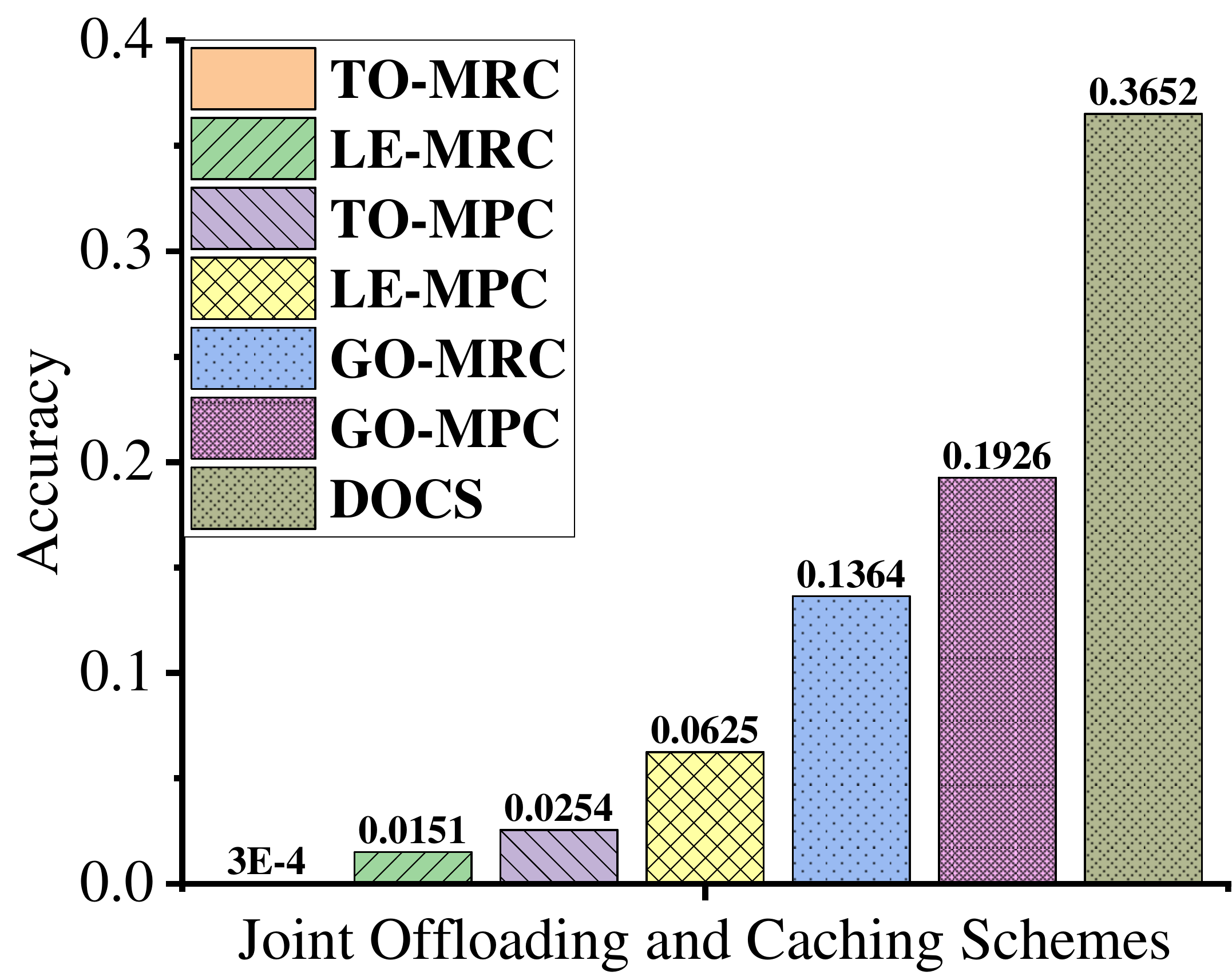}
    \caption{The accuracy of offloading and caching actions with respect to the optimal actions.}
    \label{fig:accuracy}
\end{figure}

To evaluate the performance of our proposed DOCS, we future test the accuracy for the joint offloading and caching actions with respect to the optimal actions (i.e,. the demonstrations of DIL).
Fig.~\ref{fig:accuracy} illustrates the accuracy performance.
Note that our DOCS has the highest accuracy, i.e., $36.52\%$.
The reason is that our DOCS imitates the optimal actions through offline training, thus outperforms other benchmark offloading schemes that without learning process.

\begin{figure}[t!]
    \centering
    \includegraphics[width=3in]{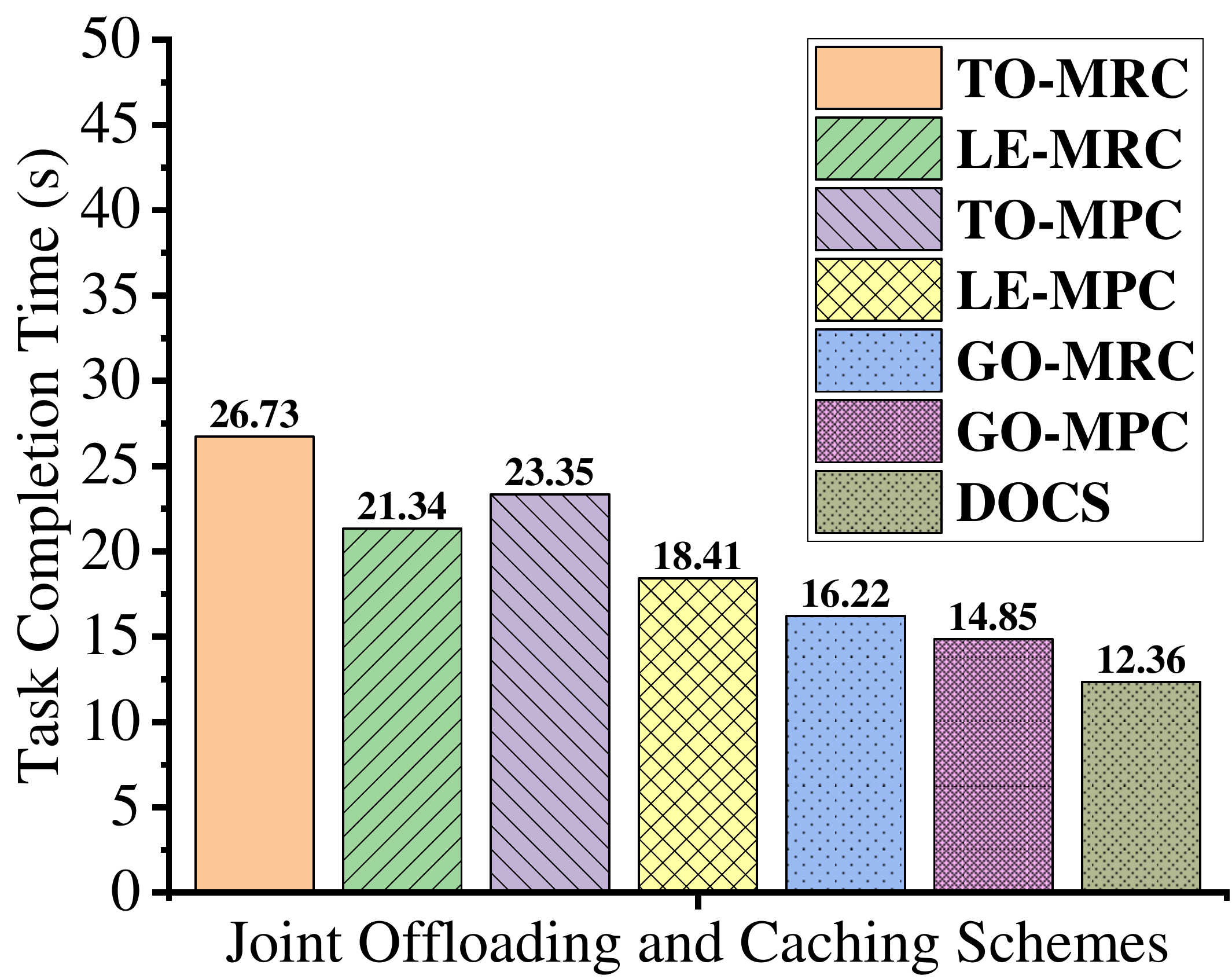}
    \caption{Task completion time for different offloading and caching schemes.}
    \label{fig:delay}
\end{figure}

Fig.~\ref{fig:delay} reports the task completion time for different offloading and caching schemes.
We can find that our DOCS can reduce at most $53.75\%$ task completion time when compared with total ofloading schemes.

\begin{figure}[t!]
    \centering
    \includegraphics[width=3in]{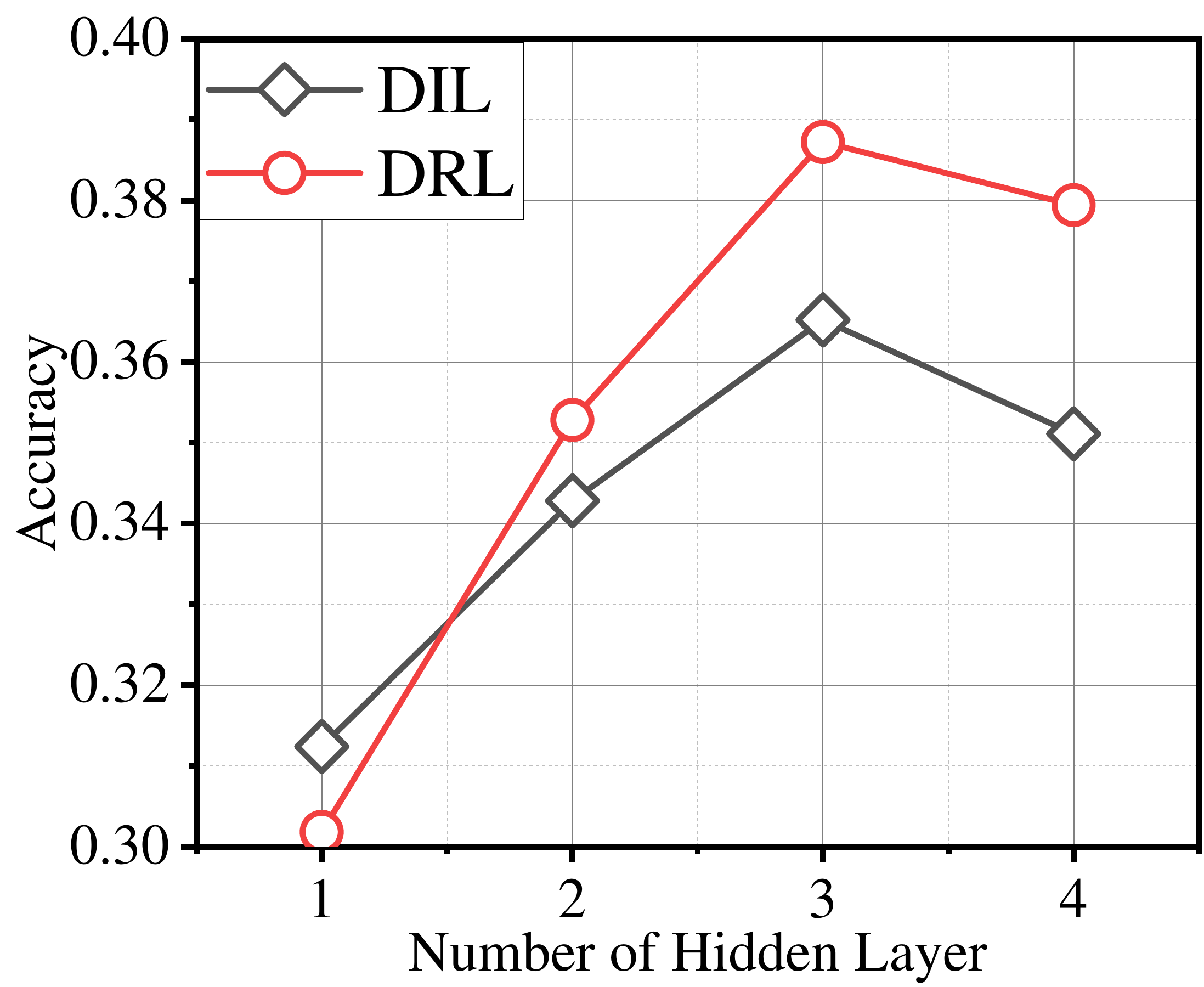}
    \caption{Accuracy performance under different hidden layer of DNN.}
    \label{fig:DIL-DRL}
\end{figure}

Fig.~\ref{fig:DIL-DRL} illustrates the accuracy performance under different number of hidden layer, where DIL represents our proposed deep imitation learning-based offloading and caching scheme.
DRL denotes the model is trained by deep reinforcement learning approach.
Note that the accuracy of both schemes increase as the number of hidden layer grows, and reach to a peak accuracy when the number is 3.
Indeed, the DRL scheme operates in an online training manner, it requires the agent to train the model continuously.
However, our DIL scheme operates in an offline-online manner.
It trains the model at a time based on historical data in offline phase, and makes real-time offloading and caching decisions in online phase.
Thus, the computation and energy consuming model training phase can be offloaded to a more powerful device (e.g., ground station in our EC-SAGINs framework).

\begin{figure}[t!]
    \centering
    \includegraphics[width=3in]{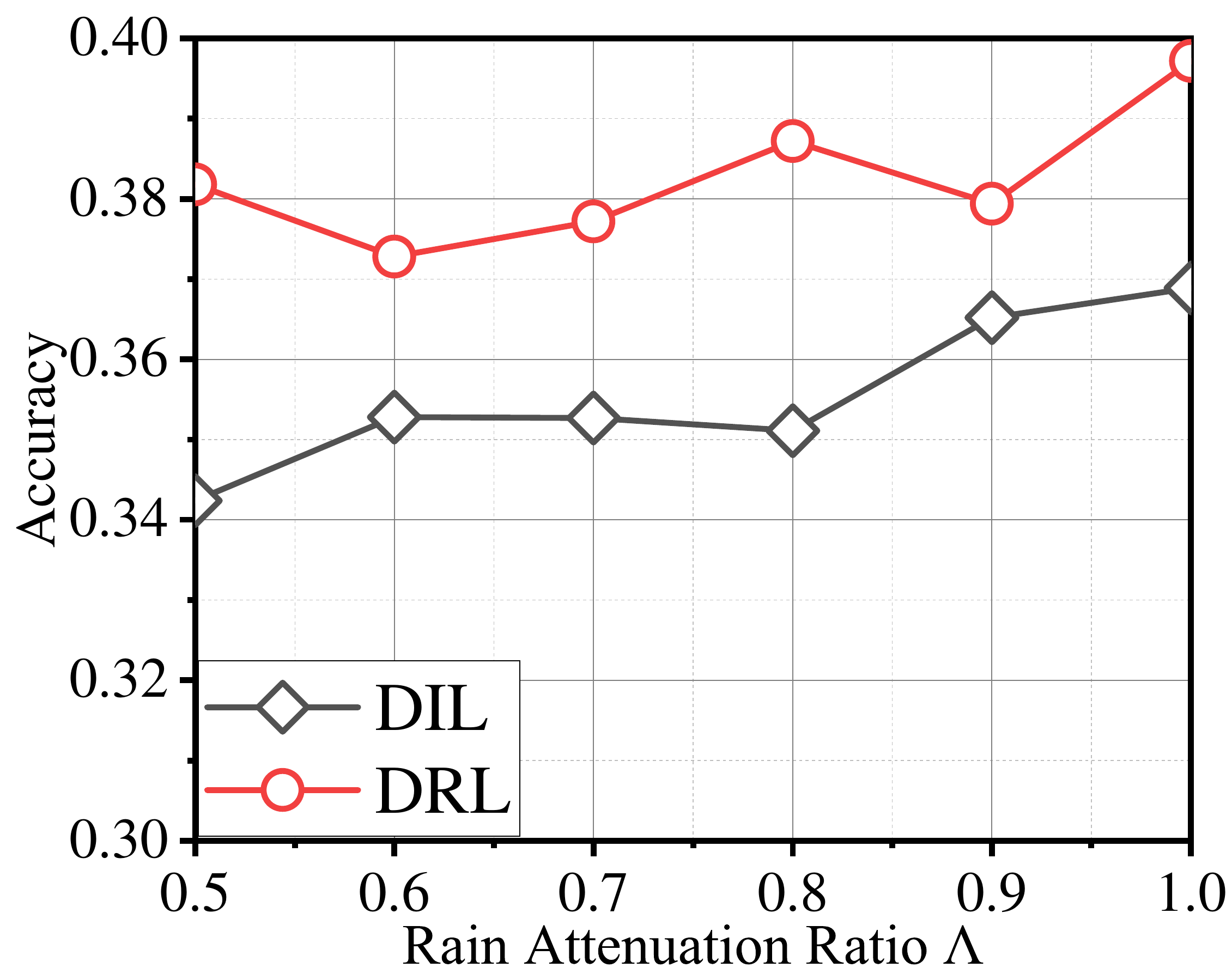}
    \caption{Accuracy performance under different rain attenuation ratio.}
    \label{fig:rain}
\end{figure}

Fig.~\ref{fig:rain} reports the accuracy performance under rain attenuation ratio.
We can find that the rain attenuation ratio has little effect on the accuracy.
It means that accuracy performance of EC-SAGINs is robust to the rain attenuation ratio (i.e., the weather between a LEO satellite and a ground station).

Based on the above simulation results, we can find that our proposed DOCS outperforms other benchmark related policies in both delay saving and decision making.
On the other hand, the task completion time for all the schemes suffer from unstable wireless link, due to the long propagation latencies of GEO relay satellite.

\section{Future Directions}{\label{sec:extensions}}
In the sections above, we focus on the edge computing techniques for SAGINs.
In this section, we further discuss some potential directions for how to efficiently integrate edge computing into SAGINs, including artificial intelligence (AI), resource allocation, security and privacy, hardware design, respectively.

\subsection{Artificial Intelligence}{\label{sec:future_AI}}
Since SAGINs have hierarchical network structures, integrating edge computing into SAGINs is more complex than the terrestrial edge computing integration.
Thus, novel and efficient techniques should be considered to optimize the network performance.
Artificial intelligence (AI) is considered to be a state-of-the-art tool to improve the performance of edge computing in SAGINs.
Compared with traditional optimization algorithm, artificial intelligence methods (e.g., deep learning) carry multiple advantages, such as i) high decision making accuracy, ii) fast inference speed, and iii) rapid and easy deployment.
In recent years, researchers have proposed a amount of research to improve network performance by leveraging AI methods, such as intelligent edge learning, intelligent resource allocation and intelligent computation offloading.
Moreover, AI methods have already been utilized in orbital network for Earth observation applications, such as image processing.

Although AI has the above advantages, it consumes more computation and communication resources than conventional methods, especially in the energy limited orbital edge computing environment.
Assume that a LEO satellite is running a deep learning algorithm, it makes sense to offload the computation intensive part (e.g., offline training) to a ground station, and keep the delay intensive part (e.g., online decision making) locally on satellite.

\subsection{Resource Allocation}{\label{sec:future_resource}}
Due to the highly dynamic network environment and complex multi-dimensional resource of SAGINs, it is necessary to design real-time and energy-efficient resource allocation schemes.
Compared to terrestrial edge computing, SAGINs have novel characteristics, such as the predicted trajectory of LEO satellites and the controllable movement of UAVs.
Thus, it makes sense to dynamically monitor the status of edge computing nodes (e.g., satellites and UAVs), including battery level, location and speed and available storage capacities.
On the other hand, limited energy supply is a key challenge of satellites/UAVs as edge servers.
Much effort must be made to develop real-time and low-cost optimization algorithm, such as deep learning based resource allocation.

\subsection{Security and Privacy}{\label{sec:future_security}}
In the future, edge computing enabled SAGINs may integrate multiple military and smart cities systems.
Thus, a large amount of privacy intensive data must be transmitted, cached and processed in a secure manner.
However, due to the open links, high mobility and dynamics of EC-SAGINs, it still faces various security threats, including message tampering, malicious attacking and jamming.
For example, it is hard for EC-SAGINs to address jamming, because of its global coverage.
In addition, traditional privacy-preserving schemes (e.g., encryption) usually result in additional delay and energy overhead, which may hamper the performance of real-time applications and energy-limited edge node.
To better protect the security and privacy of EC-SAGINs, secure transmission solutions and low-overhead privacy-preserving schemes should be investigated.

\subsection{Hardware Design}{\label{sec:future_hareware}}
To integrate the edge computing into SAGINs, we need to make some modifications on current hardware.
Since radiation is the biggest threat to orbital edge computing, how to improve radiation protection must be investigated.
Moreover, it is widely accepted that deep learning technique will plays a key role in future network, including network performance optimization and AI-based applications.
Thus it makes sense to equip enough GPU resource in edge nodes (e.g., satellites and UAVs), since GPU is more suitable for deep learning computation.
In addition, hardware expense should also be considered, especially for large scale satellite system.

\section{Conclusion}{\label{sec:Conclusion}}
In this article, we study the edge computing technologies for space-air-ground integrated networks (SAGINs) to support various IoV services.
In particular, we first review state-of-the-art orbital edge computing and aerial edge computing techniques, respectively, and discuss their architectures, use cases, advantages and challenges.
Then, we present the framework of our edge computing enabled space-air-ground integrated networks (EC-SAGINs), and elaborate the joint offloading decision and caching placement for the vehicles in remote areas.
Numerical results confirm that our proposed pre-classification scheme and DIL-based decision making algorithm outperform other benchmark policies.
At last, we future discuss some potential directions and opportunities of applying the edge computing technologies to SAGINs.

\section*{Acknowledgment}

This work was supported in part by
the National Science Foundation of China (No. 61972432, No. 61976234, No. 62002397 and No. 62072332);
the Fundamental Research Funds for the Central Universities under grant 20lgpy135;
the Guangdong Basic and Applied Basic Research Foundation(No. 2019A1515010030);
the startup fund and Intramural Grants Program 190599 provided by Auburn University;
the National Key Research and Development Program of China under Grant 2019YFB2101901.
%the National Key Research and Development Program of China under grant (No.2017YFB1001703);
%the Program for Guangdong Introducing Innovative and Entrepreneurial Teams (No.2017ZT07X355);
%the Pearl River Talent Recruitment Program (No.2017GC010465);
%the Science and Technology Program of Guangzhou under Grant 202007040006;
%the Guangdong Special Support Program under Grant 2017TX04X148.

% or


%\appendices

%\section{}

%\section*{Acknowledgment}


\ifCLASSOPTIONcaptionsoff
  \newpage
\fi

\bibliographystyle{IEEEtran}
\bibliography{IEEEabrv,bibThesis}

\end{document}